\documentclass[9pt]{osa-supplemental-document}
\setboolean{shortarticle}{false}

\newtheorem{theorem}{Theorem}

\newtheorem{proposition}[theorem]{Proposition}

\title{TPMS2STEP Supplements: Constraints Matrices and Convergence Proof of TPMS2STEP}
\author{Yaonaiming Zhao, Qiang Zou}

\begin{abstract}
Triply periodic minimal surface (TPMS) is emerging as an important way of designing microstructures. However, there has been limited use of commercial CAD/CAM/CAE software packages for TPMS design and manufacturing. This is mainly because TPMS is consistently described in the functional representation (F-rep) format, while modern CAD/CAM/CAE tools are built upon the boundary representation (B-rep) format. One possible solution to this gap is translating TPMS to STEP, which is the standard data exchange format of CAD/CAM/CAE. Following this direction, this paper proposes a new translation method with error-controlling and $C^2$ continuity-preserving features. It is based on an approximation error-driven TPMS sampling algorithm and a constrained-PIA algorithm. The sampling algorithm controls the deviation between the original and translated models. With it, an error bound of $2\epsilon$ on the deviation can be ensured if two conditions called $\epsilon$-density and $\epsilon$-approximation are satisfied. The constrained-PIA algorithm enforces $C^2$ continuity constraints during TPMS approximation, and meanwhile attaining high efficiency. A theoretical convergence proof of this algorithm is also given. The effectiveness of the translation method has been demonstrated by a series of examples and comparisons. All relevant Refs.: \cite{2005_Lin_pia_for_NURBS,2022_wang_tpms_property_porous,2021_ding_tpms_slicing,2023_careri_tpms_application_aerospace,2014_yang_tissue_engineering,2019_catchpole_tpms_application_energy_thermal_conductivity,liu2021memory,2023_hong_tpms_functional_representation,zou2023variational,2022_distance-field-rep_TPMS,2023_neural-implicit-rep_TPMS,2011_TPMS-to-mesh_visualization,2021_TPMS-to-mesh_analysis,zou2019push,zou2022robust,wang2023quasi,zou2014iso,2016_xiao_tspline_data_exchange,zou2020decision,zou2019variational,su2020accurate,li2023xvoxel,martins2021engineering,zou2013iso,zou2021length,zou2021robust,luo2023simple,wang2023computing,2022_wu_NURBS_advantages,2019_fang_tpms_definition,2021_hu_tpms_modeling_implicit, 2023_liane_tpms_modeling_boundary_curve_based,2023_jiang_meshless_tpms_based_analysis,2019_feng_tpms_mesh_visualization,2021_asbai_tpms_to_mesh,2022_kestutis_TPMS_mesh,2024_Na_TPMS_meshing,2019_savio_subdivision_surface_TPMS,2021_mesh_subdivision_tpms,2011_Pan_minimal-subdivision-surface,2015_aubry_nurbs_application,2020_noruzi_nurbs_application,1995_NURBS_BOOK,2013_fang_chord_length,1989_lee_centripetal_parameterization,1999_lim_universal_parameterization,1989_foley_knot,2016_iglesias_parameterization_optimize,2019_luo_knot_calulation, 2021_michel_heuristic_knot_placement,2005_li_adaptive_knot_placement,2015_kang_knot_placement_sparse_optimization,1992_Farin_Curvature_based_weight_assignment,2018_lin_survey_pia,2008_Nils_gauss_newton_based_iterative_solving_linear_systems,2000_Gandy_TPMS_explicit_representation_Weierstrass_Gyroid,2000_Gandy_TPMS_explicit_representation_Weierstrass_SchwarzP,2004_mit_course_chapter_18,1987_Filip_error_estimation,2000_Zheng_error_estimation_nurbs,2021_Flores_TPMS_NURBS_generation_Gyroid,1995_Gandy_TPMS_explicit_representation_Weierstrass_Diamond,2017_Hu_G2_continuity_control_points_constraints,Fogden1992,2012_kineri_surface_fitting_iterative,2024_bo_iterative_approximation_error_control_parameter_correction}
\end{abstract}

\setboolean{displaycopyright}{false} 

\begin{document}

\maketitle

\section{Part of constraint matrices for Gyroid, Diamond, and Schwarz\_P in CPIA}
\label{sec:part1}

The constraint matrices for Gyroid, Diamond, and Schwarz\_P that are not given in the article are given here.

For Gyroid, matrices $\mathbf{T}_{1g}$-$\mathbf{T}_{7g}$ and $\mathbf{M}_{ijg} (i=1,2,3,4,5$ and $ j=1,2,3,4)$ are given.
\begin{align*}
    \mathbf{T}_{1g}=\begin{bmatrix}
    0 & 0& -1 & 1 \\
    1 & 0 & 0 & 0 \\
    0 & 1 & 0 & 0 \\
    0 & 0 & 0 & 1 \\
    \end{bmatrix}
\end{align*}
\begin{align*}
    \mathbf{T}_{2g}=\begin{bmatrix}
    0 & 0 & 1 & 1 \\
    1 & 0 & 0 & -1 \\
    0 & -1 & 0 & 1 \\
    0 & 0 & 0 & 1 \\
    \end{bmatrix}
\end{align*}
\begin{align*}
    \mathbf{T}_{3g}=\begin{bmatrix}
    0 & -1 & 0 & 1.5 \\
    1 & 0 & 0 & -0.5 \\
    0 & 0 & -1 & 0.5 \\
    0 & 0 & 0 & 1 \\
    \end{bmatrix}
\end{align*}
\begin{align*}
    & \mathbf{T}_{4g}=\mathbf{T}_{1g}\mathbf{T}_{3g} \\
    & \mathbf{T}_{5g}=\mathbf{T}_{4g}^{-1} \\
    & \mathbf{T}_{6g}=\mathbf{T}_{2g}\mathbf{T}_{3g} \\
    & \mathbf{T}_{7g}=\mathbf{T}_{6g}^{-1} \\
\end{align*}
The matrices $ \mathbf{M}_{ijg} (i=1,2,3,4,5$ and $ j=1,2,3,4)$ are derived from the method mentioned below. To extract the control points and calculate the first-order and second-order derivatives, three steps are conducted: (1) number the control points; (2) map a specific part of the control points to the corresponding control points (e.g., to calculate the first-order derivatives, the second row of control points are mapped to the first row of control points); and (3) utilize the coordinates of these points to calculate the local first-order and second-order derivatives. In Step 2, the mapping is constructed using a matrix, denoted as $\mathbf{M}$. $\mathbf{M}$ is actually one of $ \mathbf{M}_{ijg} (i=1,2,3,4,5$ and $ j=1,2,3,4)$. Specifically, if the matrix $\mathbf{P}$ is the control points, then $\mathbf{MP}$ sets some rows in $\mathbf{P}$ to 0 and changes the position of other rows of elements. In the matrix $\mathbf{MP}$, a non-zero element at Row i and Column j in $\mathbf{M}$ repositions the jth line of $\mathbf{P}$ to the ith line of $\mathbf{P}$. That is, this element maps the jth control point to the ith control point. After the mapping, calculations can be done to the control points that are mapped together.

$ \mathbf{M}_{ijg} (i=1,2,3,4,5$ and $ j=1,2,3,4)$ are given as:
\begin{align*}
    &\mathbf{M}_{11g} = \sum_{i=0}^{n-5} \mathbf{J}_{n^2-n+2+i,n^2-2n+2+i} \\
    & \mathbf{M}_{21g} = \sum_{i=0}^{n-5} \mathbf{J}_{n^2-2n+2+i,n^2-2n+2+i} \\
    & \mathbf{M}_{31g} = \sum_{i=0}^{n-7} \mathbf{J}_{n^2-n+3+i,n^2-3n+3+i} \\
    & \mathbf{M}_{41g} = \sum_{i=0}^{n-7} \mathbf{J}_{n^2-2n+3+i,n^2-3n+3+i} \\
    & \mathbf{M}_{51g} = \sum_{i=0}^{n-7} \mathbf{J}_{n^2-3n+3+i,n^2-3n+3+i} 
\end{align*}
where $ \mathbf{J}_{i,j}$ is the matrix with a unique non-zero element $ 1$ at the position $ (i,j)$ and n is the number of control points of the approximation surface in the u and v direction.
\begin{align*}
    & \mathbf{M}_{12g} = \sum_{i=0}^{n-5} \mathbf{J}_{2n+in,2n+1+in} \\
    & \mathbf{M}_{22g} = \sum_{i=0}^{n-5} \mathbf{J}_{2n+1+in,2n+1+in} \\
    & \mathbf{M}_{32g} = \sum_{i=0}^{n-7} \mathbf{J}_{3n+in,3n+2+in} \\
    & \mathbf{M}_{42g} = \sum_{i=0}^{n-7} \mathbf{J}_{3n+1+in,3n+2+in} \\
    & \mathbf{M}_{52g} = \sum_{i=0}^{n-7} \mathbf{J}_{3n+2+in,3n+2+in}
\end{align*}
\begin{align*}
    & \mathbf{M}_{13g} = \sum_{i=0}^{n-3} \mathbf{J}_{2+i,n+2+i} \\
    & \mathbf{M}_{23g} = \sum_{i=0}^{n-5} \mathbf{J}_{n+2+i,n+2+i} \\
    & \mathbf{M}_{33g} = \sum_{i=0}^{n-7} \mathbf{J}_{3+i,2n+3+i} \\
    & \mathbf{M}_{43g} = \sum_{i=0}^{n-7} \mathbf{J}_{n+3+i,2n+3+i} \\
    & \mathbf{M}_{53g} = \sum_{i=0}^{n-7} \mathbf{J}_{2n+3+i,2n+3+i}
\end{align*}
\begin{align*}
    & \mathbf{M}_{14g} = \sum_{i=0}^{n-5} \mathbf{J}_{3n-1+in,3n-2+in} \\
    & \mathbf{M}_{24g} = \sum_{i=0}^{n-5} \mathbf{J}_{3n-2+in,3n-2+in} \\
    & \mathbf{M}_{34g} = \sum_{i=0}^{n-7} \mathbf{J}_{4n-1+in,4n-3+in} \\
    & \mathbf{M}_{44g} = \sum_{i=0}^{n-7} \mathbf{J}_{4n-2+in,4n-3+in} \\
    & \mathbf{M}_{54g} = \sum_{i=0}^{n-7} \mathbf{J}_{4n-3+in,4n-3+in}
\end{align*}

For Diamond, matrices $\mathbf{T}_{1d}$, $\mathbf{T}_{2d}$, $\mathbf{N}_1$, and $\mathbf{N}_2$ are given. $\mathbf{M}_{ijd}$ $(i=1,2,3,4,5$ and $ j=1,2,3,4)$ are the same as $\mathbf{M}_{ijg}$ $(i=1,2,3,4,5$ and $ j=1,2,3,4)$ mentioned before, respectively.
\begin{align*}
\mathbf{T}_{1d}=\begin{bmatrix}
 0   & -1 & 0 & 0\\
 1  &  0 & 0 & 0  \\
 0 & 0 & -1 & 0 \\
 0 & 0 & 0 & 1
\end{bmatrix}
\end{align*}
\begin{align*}
\mathbf{T}_{2d}=\begin{bmatrix}
  -\frac{1}{2}   & -\frac{1}{2} & \frac{\sqrt[]{2} }{2} & \frac{\sqrt[]{2} }{2}\\
 -\frac{1}{2}  &  -\frac{1}{2} &  -\frac{\sqrt[]{2} }{2} & \frac{\sqrt[]{2} }{2}  \\
 \frac{\sqrt[]{2} }{2} & -\frac{\sqrt[]{2} }{2} &0  & 0 \\
 0 & 0 & 0 & 1
\end{bmatrix}
\end{align*}
\begin{align*}
    \mathbf{N}_1&=\begin{bmatrix}
 \mathbf{I}_{n^2\times n^2}  &\mathbf{0}_{n^2\times n^2}  \\
 \mathbf{0}_{n^2\times n^2}  & \mathbf{0}_{n^2\times n^2} 
\end{bmatrix}_{2n^2\times 2n^2} \\
\mathbf{N}_2& =\begin{bmatrix}
 \mathbf{0}_{n^2\times n^2}  & \mathbf{I}_{n^2\times n^2} \\
 \mathbf{0}_{n^2\times n^2}  & \mathbf{0}_{n^2\times n^2} 
\end{bmatrix}_{2n^2\times 2n^2}
\end{align*}
where n is the number of control points of the approximation surface in the u and v direction. $\mathbf{I}$ is the identity matrix. $\mathbf{N}_1$ and $\mathbf{N}_2$ are designed to satisfy the equations $ \mathbf{N}_1\mathbf{P}=\begin{pmatrix}
 \mathbf{P}_1 \\
\mathbf{0}
\end{pmatrix}$ and $ \mathbf{N}_2\mathbf{P}=\begin{pmatrix}
 \mathbf{P}_2 \\
\mathbf{0}
\end{pmatrix}$ where $ \mathbf{P}=\begin{pmatrix}
 \mathbf{P}_1 \\
\mathbf{P}_2
\end{pmatrix}$.

For Schwarz\_P, matrices $ \mathbf{T}_{1p}$-$ \mathbf{T}_{4p}$ are given. $\mathbf{M}_{ijp}$ $(i=1,2,3,4,5$ and $ j=1,2,3,4)$ are the same as $\mathbf{M}_{ijg}$ $(i=1,2,3,4,5$ and $ j=1,2,3,4)$ mentioned above.

\begin{align*}
\mathbf{T}_{1p}=\begin{bmatrix}
 \frac{1}{2} & -\frac{1}{2} & -\frac{\sqrt[]{2}}{2} & -\frac{\sqrt[]{2}}{4} \\
 -\frac{1}{2} & \frac{1}{2} & -\frac{\sqrt[]{2}}{2} & -\frac{\sqrt[]{2}}{4} \\
 -\frac{\sqrt[]{2}}{2} & -\frac{\sqrt[]{2}}{2} & 0 & -\frac{1}{2} \\
 0 & 0 & 0 & 1
\end{bmatrix}
\end{align*}
\begin{align*}
\mathbf{T}_{2p}=\begin{bmatrix}
 \frac{1}{2} & \frac{1}{2} & -\frac{\sqrt[]{2}}{2} & \frac{\sqrt[]{2}}{4} \\
 \frac{1}{2} & \frac{1}{2} & \frac{\sqrt[]{2}}{2} & -\frac{\sqrt[]{2}}{4} \\
 -\frac{\sqrt[]{2}}{2} & \frac{\sqrt[]{2}}{2} & 0 & \frac{1}{2} \\
 0 & 0 & 0 & 1
\end{bmatrix}
\end{align*}
\begin{align*}
\mathbf{T}_{3p}=\begin{bmatrix}
0 & -1 &  0 & 0 \\
-1 & 0  &  0 & 0 \\
0 & 0 &  1 & 0 \\
 0 & 0 & 0 & 1
\end{bmatrix}
\end{align*}
\begin{align*}
\mathbf{T}_{4p}=\begin{bmatrix}
0 & 1 &  0 & 0 \\
1 & 0  &  0 & 0 \\
0 & 0 &  1 & 0 \\
 0 & 0 & 0 & 1
\end{bmatrix}
\end{align*}


\section{Convergence proofs for Diamond and Schwarz\_P TPMS}
The following sections include the convergence proofs for Diamond and Schwarz\_P TPMS.

\begin{proposition}
The CPIA iterative method for Diamond is convergent and the limit surface is the least-square fitting outcome of the initial data $\mathrm{\{\mathbf{Q}_{ij}\}^{m_1,m_2}_{i=0,j=0}}$.
\end{proposition}

\textbf{Proof.} As the result of the iterative procedure of CPIA, two sequences of the control points of the offset surface for Diamond $\mathrm{\{\mathbf{P}_1^{k}(u,v),k=0,1,\cdots \}}$ and $\mathrm{\{\mathbf{P}_2^{k}(u,v),k=0,1,\cdots \}}$ are generated. To show their convergence, let $\mathbf{P}^k=\{\mathbf{P}_0^k,\mathbf{P}_1^k,\cdots,\mathbf{P}_n^k\}^T$ and $\mathbf{Q}=\{\mathbf{Q}_0,\mathbf{Q}_1,\cdots,\mathbf{Q}_m \}^T$.
Here $ \mathbf{P}=\begin{pmatrix}
 \mathbf{P}_1\\
\mathbf{P}_2
\end{pmatrix}$ and $ \mathbf{Q}=\begin{pmatrix}
 \mathbf{Q}_1\\
\mathbf{Q}_2
\end{pmatrix}$.

In the (k+1)th iteration, we can derive the coordinates of the newly adjusted control points as:
\begin{align}\label{eq:c_d}
\begin{aligned}
    \mathbf{P}^{k+1} &= \mathbf{P}^k+(\mathbf{Q-BwP}^k)+\frac{1}{2}\sum_{i=1}^{2} ((\mathbf{M}_{24d}-\mathbf{M}_{14d})\mathbf{N}_{i} \mathbf{P}^k\mathbf{T}_{1d}-(\mathbf{M}_{21d}-\mathbf{M}_{11d})\mathbf{N}_{3-i} \mathbf{P}^k) \\
    &+\frac{1}{2}\sum_{i=1}^{2} ((3\mathbf{M}_{44d}-2\mathbf{M}_{34d}-\mathbf{M}_{54d})\mathbf{N}_{i} \mathbf{P}^k\mathbf{T}_{1d}-(3\mathbf{M}_{41d}-2\mathbf{M}_{31d} -\mathbf{M}_{51d})\mathbf{N}_{3-i} \mathbf{P}^k) \\
    &+\frac{1}{2}\sum_{i=1}^{2} ((\mathbf{M}_{23d}-\mathbf{M}_{13d})\mathbf{N}_{i} \mathbf{P}^k\mathbf{T}_{2d}-(\mathbf{M}_{22d}-\mathbf{M}_{12d})\mathbf{N}_{3-i} \mathbf{P}^k) \\
    &+\frac{1}{2}\sum_{i=1}^{2}((3\mathbf{M}_{43d}-2\mathbf{M}_{33d}-\mathbf{M}_{53d})\mathbf{N}_{i} \mathbf{P}^k\mathbf{T}_{2d}-(3\mathbf{M}_{42d}-2\mathbf{M}_{32d}-\mathbf{M}_{52d})\mathbf{N}_{3-i} \mathbf{P}^k) 
\end{aligned}
\end{align}
The matrices $\mathbf{T}_{id} (i=1,2) $ are given in Sec.~\ref{sec:part1} and they are invertible matrices where the absolute values of the eigenvalues are all 1, i.e., $\left | \lambda_i ( \mathbf{T}_{jd})\right| = 1, i=1,2,3,4$ and $ j=1,2$. $\mathbf{B}$ refers to the B-spline basis function matrix and $\mathbf{w}$ is the weight matrix. The matrices $ \mathbf{M}_{ijd}$ $ (i=1,2,3,4,5$ and $ j=1,2,3,4)$, $\mathbf{N}_1$, and $\mathbf{N}_2$ are also given in Sec.~\ref{sec:part1}.
To simplify the form of the equations, let $\alpha_i =3\mathbf{M}_{4id}-2\mathbf{M}_{3id}-\mathbf{M}_{5id}$, $\beta_i = \mathbf{M}_{2id}-\mathbf{M}_{1id} (i=1,2,3,4)$, $\mathbf{D} = \mathbf{I}-\mathbf{Bw}$, and $\mathbf{r}^k=\mathbf{P}^k-\mathbf{w}^{-1}\mathbf{B}^{-1}\mathbf{Q}$.
Then Eq.~\ref{eq:c_d} can be transformed into a new form:
\begin{align}
\begin{aligned}
    \mathbf{r}^{k+1} = & (\mathbf{D}-\frac{1}{2}(\alpha_1+\beta_1+\alpha_2+\beta_2)(\mathbf{N}_1+\mathbf{N}_2))\mathbf{r}^k +\frac{1}{2}(\alpha_4+\beta_4)(\mathbf{N}_1+\mathbf{N}_2)\mathbf{r}^k\mathbf{T}_{1d}+ \frac{1}{2}(\alpha_3+\beta_3)(\mathbf{N}_1+\mathbf{N}_2)\mathbf{r}^k\mathbf{T}_{2d}\\
    & = \cdots \\
    & = \sum_{i_1=0}^{k+1}\sum_{i_2=0}^{k+1}\sum_{i_3=0}^{k+1} { k+1\choose i_1,i_2,i_3}(\mathbf{D}-\frac{1}{2}(\alpha_1+\beta_1+\alpha_2+\beta_2)(\mathbf{N}_1+\mathbf{N}_2))^{i_1} \\
    &\left[ \frac{(\alpha_4+\beta_4)(\mathbf{N}_1+\mathbf{N}_2)}{2}\right]^{i_2} \left[\frac{(\alpha_3+\beta_3)(\mathbf{N}_1+\mathbf{N}_2)}{2}\right]^{i_3} \mathbf{r}^0\mathbf{T}_{1d}^{i_2}\mathbf{T}_{2d}^{i_3}\
\end{aligned}
\end{align}

Supposing $\{ \lambda_k(\alpha_i) \}(k=0,1,\cdots,n-1)$ and $\{ \lambda_{l}(\beta_j) \}(l=0,1,\cdots,n-1)$ are the eigenvalues of $ \alpha_i$ and $ \beta_j$ sorted in non-decreasing order. n is the number of control points in u and v direction.
Since $\mid \lambda_k(\alpha_i) \mid=0\quad or \quad 1,i=1,2,3,4,k=0,1,\cdots,n-1$, $ \mid \lambda_l(\beta_j) \mid=0\quad or \quad 1,j=1,2,3,4,l=0,1,\cdots,n-1$, $ r(\sum_{i=0}^{4}\alpha_i)<n$, $ r(\sum_{i=0}^{4}\beta_i)<n$, $\left | \lambda_i ( \mathbf{T}_{jd})\right| = 1, i=1,2,3,4, j=1,2$, and $ r(\mathbf{N}_1+\mathbf{N}_2)=\frac{n}{2}<n$, their powers have no effect on convergence. $ r(\alpha)$ is the rank of $ \alpha$.

Let
\begin{align}
    \mathbf{s}^{k+1}=\mathbf{D}^{k+1} \mathbf{r}^0
\end{align}
Similar to Gyroid, we first show that $\mathbf{s}^{k+1}$ is convergent. Then we show that $\mathbf{r}^{k+1}$ is also convergent. To achieve this goal, $\mathbf{s}^{k+1} $ is turned into the form
\begin{align}
\mathbf{s}^{k+1}=(\mathbf{I}-\mathbf{Bw})^{k+1} \mathbf{r}^0
\end{align}
Since \textbf{B} is a non-singular matrix, it is positive definite. From Theorem 2.2 in ~\cite{2005_Lin_pia_for_NURBS}, we know that $\rho (\mathbf{I}-\mathbf{B})<1$, where $\rho (\mathbf{I}-\mathbf{B})$ is the spectral radius of $\mathbf{I}-\mathbf{B}$. With uniform weight assignment, $\rho (\mathbf{I}-\mathbf{Bw})<1$. Therefore, we get the following equation:
\begin{equation}
\lim_{k \to \infty} (\mathbf{I}-\mathbf{Bw})^k= \lim_{k \to \infty} \mathbf{s}^{k+1} =(\mathbf{0})_{n+1}
\end{equation}
Then each term in $\mathbf{r}^{k+1}$ has the same spectral radius as each term in $\mathbf{s}^{k+1}$, as is proved for Gyroid. So if $\mathbf{s}^{k+1}$ is convergent, $\mathbf{r}^{k+1}$ is also convergent, which could be expressed as:
\begin{equation}
\lim_{k \to \infty} \mathbf{P}^k-\mathbf{w}^{-1}\mathbf{B}^{-1}\mathbf{Q}=\mathbf{0}
\end{equation}
So \{ $\mathbf{P}^k$ \} is convergent, and
\begin{equation}
\mathbf{P}^ \infty=\mathbf{w}^{-1}\mathbf{B}^{-1}\mathbf{Q}
\end{equation}

\begin{proposition}
The CPIA iterative method for Schwarz\_P is convergent and the limit surface is the least-square fitting outcome of the initial data $\mathrm{\{\mathbf{Q}_{ij}\}^{m_1,m_2}_{i=0,j=0}}$.
\end{proposition}

\textbf{Proof.} As the result of the iterative procedure of CPIA, a sequence of the control points of the offset surface for Schwarz\_P $\mathrm{\{\mathbf{P}^{k}(u,v),k=0,1,\cdots \}}$ is generated. To show its convergence, let $\mathbf{P}^k=\mathbf{P}_0^k,\mathbf{P}_1^k,\cdots,\mathbf{P}_n^k\}^T$ and $\mathbf{Q}=\{\mathbf{Q}_0,\mathbf{Q}_1,\cdots,\mathbf{Q}_m \}^T$.
In the (k+1)th iteration, we have
\begin{align}\label{eq:c_p}
\begin{aligned}
    \mathbf{P}^{k+1} &= \mathbf{P}^k+(\mathbf{Q-BwP}^k)+\frac{1}{2} \sum_{i=1}^{4} ((\mathbf{M}_{2ip}-\mathbf{M}_{1ip})\mathbf{P}^k\mathbf{T}_{ip}-(\mathbf{M}_{2ip}-\mathbf{M}_{1ip})\mathbf{P}^k) \\
     &+\frac{1}{2}\sum_{i=1}^{4}((3\mathbf{M}_{4ip}-2\mathbf{M}_{3ip}-\mathbf{M}_{5ip})\mathbf{P}^k\mathbf{T}_{ip}-(3\mathbf{M}_{4ip}-2\mathbf{M}_{3ip}-\mathbf{M}_{5ip})\mathbf{P}^k)
     \end{aligned}
\end{align}
where $\mathbf{M}_{ijp} (i=1,2,3,4,5, j=1,2,3,4)$ and $\mathbf{T}_{ip} (i=1,2,3,4)$ are given in Sec.~\ref{sec:part1}. $\mathbf{T}_{ip} (i=1,2,3,4) $ are invertible matrices where the absolute values of the eigenvalues are all 1, i.e., $\left | \lambda_i ( \mathbf{T}_{jd})\right| = 1, i=1,2,3,4$ and $ j=1,2,3,4$. $\mathbf{B}$ and $\mathbf{w}$ are mentioned in the proof for Gyroid and Diamond.

To simplify the expression of the formulas, 
let $\alpha_i =3\mathbf{M}_{4ip}-2\mathbf{M}_{3ip}-\mathbf{M}_{5ip}$ and $\beta_i = \mathbf{M}_{2ip}-\mathbf{M}_{1ip} (i=1,2,3,4)$.

Let $ \mathbf{D} = \mathbf{I}-\mathbf{Bw}$ and $\mathbf{r}^k=\mathbf{P}^k-\mathbf{w}^{-1}\mathbf{B}^{-1}\mathbf{Q}$.
Then a new form of Eq.~\ref{eq:c_p} could be derived as
\begin{align}
\begin{aligned}
    \mathbf{r}^{k+1} = & (\mathbf{D}-\frac{1}{2}\sum_{i=0}^{4}(\alpha_i+\beta_i))\mathbf{r}^k+ \frac{1}{2} \sum_{i=0}^{4}(\alpha_i+\beta_i) \mathbf{r}^k\mathbf{T}_{ip}\\
    & = \cdots \\
    & = \sum_{i=0}^{k+1} { k+1\choose i}(\mathbf{D}-\frac{1}{2}\sum_{n=0}^{4}(\alpha_n+\beta_n))^{i}(\frac{1}{2}\sum_{n=0}^{4}(\alpha_n+\beta_n))^{k+1-i}\mathbf{r}^0\mathbf{T}_{ip}^{k+1-i}
\end{aligned}
\end{align}
Supposing $\{ \lambda_k(\alpha_i) \}(k=0,1,\cdots,n-1)$ and $\{ \lambda_l(\beta_j) \}(l=0,1,\cdots,n-1)$ are the eigenvalues of $ \alpha_i$ and $ \beta_i$ sorted in non-decreasing order. Since $\mid \lambda_k(\alpha_i) \mid=0\quad or \quad 1,k=0,1,\cdots,n-1$, $\mid \lambda_l(\beta_j) \mid=0\quad or \quad 1,l=0,1,\cdots,n-1$, and $ r(\sum_{i=0}^{4}\beta_i)<n$, their powers have no effect on convergence. ($r(\sum_{i=0}^{4}\beta_i)$ is the rank of $ (\sum_{i=0}^{4}\beta_i)$). 

Let
\begin{equation}
    \mathbf{s}^{k+1}=\mathbf{D}^{k+1}\mathbf{r}^0
\end{equation}
Then the convergence of $\mathbf{r}^{k+1}$ is the same as $\mathbf{s}^{k+1}$. Then we only need to show that $\{ \mathbf{s}^{k}\}$ is convergent. $ \mathbf{s}^{k+1}$ have another form
\begin{align}
\mathbf{s}^{k+1}=(\mathbf{I}-\mathbf{Bw})^{k+1} \mathbf{r}^0
\end{align}
Since \textbf{B} is a non-singular matrix, it is positive definite. From Theorem 2.2 in ~\cite{2005_Lin_pia_for_NURBS}, we know that $\rho (\mathbf{I}-\mathbf{B})<1$, where $\rho (\mathbf{I}-\mathbf{B})$ is the spectral radius of $\mathbf{I}-\mathbf{B}$. With uniform weight assignment, $\rho (\mathbf{I}-\mathbf{Bw})<1$. Therefore, we have the following equation
\begin{equation}
    \lim_{k \to \infty} (\mathbf{I}-\mathbf{Bw})^k= \lim_{k \to \infty} \mathbf{s}^{k+1} =(\mathbf{0})_{n+1}
\end{equation}
After proving the convergence of $\mathbf{s}^{k+1}$, $\mathbf{r}^{k+1}$ is also convergent according to the proof for Gyroid. This convergence has the form:
\begin{equation}
\lim_{k \to \infty} \mathbf{P}^k-\mathbf{w}^{-1}\mathbf{B}^{-1}\mathbf{Q}=\mathbf{0}
\end{equation}
So \{ $\mathbf{P}^k$ \} is convergent, and
\begin{equation}
\mathbf{P}^ \infty=\mathbf{w}^{-1}\mathbf{B}^{-1}\mathbf{Q}
\end{equation}




\section{THE SECOND ORDER DERIVATIVES OF THE OFFSET EQUATION}
\label{sec:part3}
Given $ \phi_1$, $ \phi_2$, and $ \phi_3$ in Section 3.1 of the paper, the real and imaginary parts of the three complex variables, denoted as $ \mathbf{p}_1 \mathbf{p}_2 \mathbf{p}_3$ and $ \mathbf{q}_1 \mathbf{q}_2 \mathbf{q}_3$, are given as:
\begin{align}
    \begin{aligned}
        & \mathbf{p}_1=Re(\phi_1) \\
        & \mathbf{p}_2=Re(\phi_2) \\
        & \mathbf{p}_3=Re(\phi_3) \\
        & \mathbf{q}_1=Im(\phi_1) \\
        & \mathbf{q}_2=Im(\phi_2) \\
        & \mathbf{q}_3=Im(\phi_3) \\
    \end{aligned}
\end{align}
To calculate the second order derivatives, the first order and second order derivatives of $ \phi_1$, $ \phi_2$, and $ \phi_3$ are first calculated, given as:
\begin{small}
\begin{align}
    \begin{aligned}
        & \mathbf{p}^{\prime }_1=Re\left[-\frac{\left(1-\tau^2\right) \left(8 \tau^7-56 \tau^3\right)}{2 \left(\tau^8-14 \tau^4+1\right)^{3/2}}-\frac{2 \tau}{\sqrt{\tau^8-14 \tau^4+1}}\right] \\
        & \mathbf{p}^{\prime }_2=Re\left[\frac{2 i \tau}{\sqrt{\tau^8-14 \tau^4+1}}-\frac{i \left(\tau^2+1\right) \left(8 \tau^7-56 \tau^3\right)}{2 \left(\tau^8-14 \tau^4+1\right)^{3/2}}\right] \\
        & \mathbf{p}^{\prime }_3=Re\left[\frac{2}{\sqrt{\tau^8-14 \tau^4+1}}-\frac{\tau \left(8 \tau^7-56 \tau^3\right)}{\left(\tau^8-14 \tau^4+1\right)^{3/2}}\right] \\
        & \mathbf{q}^{\prime }_1=Im\left[-\frac{\left(1-\tau^2\right) \left(8 \tau^7-56 \tau^3\right)}{2 \left(\tau^8-14 \tau^4+1\right)^{3/2}}-\frac{2 \tau}{\sqrt{\tau^8-14 \tau^4+1}}\right] \\
        & \mathbf{q}^{\prime }_1=Im\left[\frac{2 i \tau}{\sqrt{\tau^8-14 \tau^4+1}}-\frac{i \left(\tau^2+1\right) \left(8 \tau^7-56 \tau^3\right)}{2 \left(\tau^8-14 \tau^4+1\right)^{3/2}}\right] \\
        & \mathbf{q}^{\prime }_3=Im\left[\frac{2}{\sqrt{\tau^8-14 \tau^4+1}}-\frac{\tau \left(8 \tau^7-56 \tau^3\right)}{\left(\tau^8-14 \tau^4+1\right)^{3/2}}\right] \\
    \end{aligned}
\end{align}
\begin{align}
    \begin{aligned}
    & \mathbf{p}^{\prime\prime }_1=Re\left[-\frac{2}{\sqrt{\tau^8-14 \tau^4+1}}+\frac{2 \tau \left(8 \tau^7-56 \tau^3\right)}{\left(\tau^8-14 \tau^4+1\right)^{3/2}}+\left(1-\tau^2\right) \left(\frac{3 \left(8 \tau^7-56 \tau^3\right)^2}{4 \left(\tau^8-14 \tau^4+1\right)^{5/2}}-\frac{56 \tau^6-168 \tau^2}{2 \left(\tau^8-14 \tau^4+1\right)^{3/2}}\right)\right] \\
    & \mathbf{p}^{\prime\prime }_2=Re\left[\frac{2 i}{\sqrt{\tau^8-14 \tau^4+1}}-\frac{2 i \tau \left(8 \tau^7-56 \tau^3\right)}{\left(\tau^8-14 \tau^4+1\right)^{3/2}}+i \left(\tau^2+1\right) \left(\frac{3 \left(8 \tau^7-56 \tau^3\right)^2}{4 \left(\tau^8-14 \tau^4+1\right)^{5/2}}-\frac{56 \tau^6-168 \tau^2}{2 \left(\tau^8-14 \tau^4+1\right)^{3/2}}\right)\right] \\
    & \mathbf{p}^{\prime\prime }_3= Re\left[2 \tau \left(\frac{3 \left(8 \tau^7-56 \tau^3\right)^2}{4 \left(\tau^8-14 \tau^4+1\right)^{5/2}}-\frac{56 \tau^6-168 \tau^2}{2 \left(\tau^8-14 \tau^4+1\right)^{3/2}}\right)-\frac{2 \left(8 \tau^7-56 \tau^3\right)}{\left(\tau^8-14 \tau^4+1\right)^{3/2}}\right] \\
    & \mathbf{q}^{\prime\prime }_1=Im\left[-\frac{2}{\sqrt{\tau^8-14 \tau^4+1}}+\frac{2 \tau \left(8 \tau^7-56 \tau^3\right)}{\left(\tau^8-14 \tau^4+1\right)^{3/2}}+\left(1-\tau^2\right) \left(\frac{3 \left(8 \tau^7-56 \tau^3\right)^2}{4 \left(\tau^8-14 \tau^4+1\right)^{5/2}}-\frac{56 \tau^6-168 \tau^2}{2 \left(\tau^8-14 \tau^4+1\right)^{3/2}}\right)\right] \\
    & \mathbf{q}^{\prime\prime }_2=Im\left[\frac{2 i}{\sqrt{\tau^8-14 \tau^4+1}}-\frac{2 i \tau \left(8 \tau^7-56 \tau^3\right)}{\left(\tau^8-14 \tau^4+1\right)^{3/2}}+i \left(\tau^2+1\right) \left(\frac{3 \left(8 \tau^7-56 \tau^3\right)^2}{4 \left(\tau^8-14 \tau^4+1\right)^{5/2}}-\frac{56 \tau^6-168 \tau^2}{2 \left(\tau^8-14 \tau^4+1\right)^{3/2}}\right)\right] \\
    & \mathbf{q}^{\prime\prime }_3= Im\left[2 \tau \left(\frac{3 \left(8 \tau^7-56 \tau^3\right)^2}{4 \left(\tau^8-14 \tau^4+1\right)^{5/2}}-\frac{56 \tau^6-168 \tau^2}{2 \left(\tau^8-14 \tau^4+1\right)^{3/2}}\right)-\frac{2 \left(8 \tau^7-56 \tau^3\right)}{\left(\tau^8-14 \tau^4+1\right)^{3/2}}\right] \\
    \end{aligned}
\end{align}
\end{small}
where $\tau$ is the complex variable mentioned in Section 3.1 of the paper. $ \mathbf{p}^{\prime }_1 \mathbf{p}^{\prime }_2 \mathbf{p}^{\prime }_3$ and $ \mathbf{q}^{\prime }_1 \mathbf{q}^{\prime }_2 \mathbf{q}^{\prime }_3$ are the real and imaginary parts of $ \phi_1^{\prime} \phi_2^{\prime} \phi_3^{\prime}$. $ \mathbf{p}^{\prime\prime }_1 \mathbf{p}^{\prime\prime }_2 \mathbf{p}^{\prime\prime }_3$ and $ \mathbf{q}^{\prime\prime }_1 \mathbf{q}^{\prime\prime }_2 \mathbf{q}^{\prime\prime }_3$ are the real and imaginary parts of $ \phi_1^{\prime\prime} \phi_2^{\prime\prime} \phi_3^{\prime\prime}$.

Then the second-order derivatives of the offset equation are given as:
\begin{tiny}
\begin{align}
    \begin{aligned}
        x^{\prime\prime}=& \frac{8\mathbf{A}_1(\mathbf{q}_3\mathbf{p}^{\prime}_2+\mathbf{p}_2\mathbf{q}^{\prime}_3-\mathbf{q}_2\mathbf{p}^{\prime}_3-\mathbf{p}_3\mathbf{q}^{\prime}_2)\left[ (\mathbf{p}_2\mathbf{q}_1-\mathbf{p}_1\mathbf{q}_2)(\mathbf{q}_1\mathbf{p}_2^{\prime}+\mathbf{p}_2\mathbf{q}_1^{\prime}-\mathbf{q}_2\mathbf{p}_1^{\prime}-\mathbf{p}_1\mathbf{q}_2^{\prime})+(\mathbf{p}_3\mathbf{q}_1-\mathbf{p}_1\mathbf{q}_3)(\mathbf{q}_1\mathbf{p}_3^{\prime}+\mathbf{p}_3\mathbf{q}_1^{\prime}-\mathbf{q}_3\mathbf{p}_1^{\prime}-\mathbf{p}_1\mathbf{q}_3^{\prime})+(\mathbf{p}_3\mathbf{q}_2-\mathbf{p}_2\mathbf{q}_3)(\mathbf{q}_2\mathbf{p}_3^{\prime}+\mathbf{p}_3\mathbf{q}_2^{\prime}-\mathbf{q}_3\mathbf{p}_2^{\prime}-\mathbf{p}_2\mathbf{q}_3^{\prime}) \right] }{4\mathbf{A}_1^{\frac{5}{2}}} \\
        & +\frac{4\mathbf{A}_1^{2}(2\mathbf{p}_3^{\prime}\mathbf{q}_2^{\prime}-2\mathbf{q}_3^{\prime}\mathbf{p}_2^{\prime}-\mathbf{q}_3\mathbf{p}_2^{\prime\prime}+\mathbf{q}_2\mathbf{p}_3^{\prime\prime}+\mathbf{p}_3\mathbf{q}_2^{\prime\prime}-\mathbf{p}_2\mathbf{q}_3^{\prime\prime})}{4\mathbf{A}_1^{\frac{5}{2}}} \\
        & +\frac{12(\mathbf{p}_3\mathbf{q}_2-\mathbf{p}_2\mathbf{q}_3)\left[ (\mathbf{p}_2\mathbf{q}_1-\mathbf{p}_1\mathbf{q}_2)(\mathbf{q}_1\mathbf{p}_2^{\prime}+\mathbf{p}_2\mathbf{q}_1^{\prime}-\mathbf{q}_2\mathbf{p}_1^{\prime}-\mathbf{p}_1\mathbf{q}_2^{\prime})+(\mathbf{p}_3\mathbf{q}_1-\mathbf{p}_1\mathbf{q}_3)(\mathbf{q}_1\mathbf{p}_3^{\prime}+\mathbf{p}_3\mathbf{q}_1^{\prime}-\mathbf{q}_3\mathbf{p}_1^{\prime}-\mathbf{p}_1\mathbf{q}_3^{\prime})+(\mathbf{p}_3\mathbf{q}_2-\mathbf{p}_2\mathbf{q}_3)(\mathbf{q}_2\mathbf{p}_3^{\prime}+\mathbf{p}_3\mathbf{q}_2^{\prime}-\mathbf{q}_3\mathbf{p}_2^{\prime}-\mathbf{p}_2\mathbf{q}_3^{\prime}) \right ]^{2}}{4\mathbf{A}_1^{\frac{5}{2}}} \\
        & -\frac{4\mathbf{A}_1(\mathbf{p}_3\mathbf{q}_2-\mathbf{p}_2\mathbf{q}_3)\left[(\mathbf{q}_2\mathbf{p}_1^{\prime}+\mathbf{p}_1\mathbf{q}_2^{\prime}-\mathbf{q}_1\mathbf{p}_2^{\prime}-\mathbf{p}_2\mathbf{q}_1^{\prime})^{2}+(\mathbf{q}_3\mathbf{p}_1^{\prime}+\mathbf{p}_1\mathbf{q}_3^{\prime}-\mathbf{q}_1\mathbf{p}_3^{\prime}-\mathbf{p}_3\mathbf{q}_1^{\prime})^{2}+(\mathbf{q}_3\mathbf{p}_2^{\prime}+\mathbf{p}_2\mathbf{q}_3^{\prime}-\mathbf{q}_2\mathbf{p}_3^{\prime}-\mathbf{p}_3\mathbf{q}_2^{\prime})^{2}\right]}{4\mathbf{A}_1^{\frac{5}{2}}} \\
        & -\frac{4\mathbf{A}_1(\mathbf{p}_3\mathbf{q}_2-\mathbf{p}_2\mathbf{q}_3)\left[(\mathbf{p}_2\mathbf{q}_1-\mathbf{p}_1\mathbf{q}_2)(2\mathbf{p}_2^{\prime}\mathbf{q}_1^{\prime}-2\mathbf{p}_1^{\prime}\mathbf{q}_2^{\prime}-\mathbf{q}_2\mathbf{p}_1^{\prime\prime}+\mathbf{q}_1\mathbf{p}_2^{\prime\prime}+\mathbf{p}_2\mathbf{q}_1^{\prime\prime}-\mathbf{p}_1\mathbf{q}_2^{\prime\prime})+(\mathbf{p}_3\mathbf{q}_1-\mathbf{p}_1\mathbf{q}_3)(2\mathbf{p}_3^{\prime}\mathbf{q}_1^{\prime}-2\mathbf{p}_1^{\prime}\mathbf{q}_3^{\prime}-\mathbf{q}_3\mathbf{p}_1^{\prime\prime}+\mathbf{q}_1\mathbf{p}_3^{\prime\prime}+\mathbf{p}_3\mathbf{q}_1^{\prime\prime}-\mathbf{p}_1\mathbf{q}_3^{\prime\prime})\right]}{4\mathbf{A}_1^{\frac{5}{2}}} \\
        & -\frac{4\mathbf{A}_1(\mathbf{p}_3\mathbf{q}_2-\mathbf{p}_2\mathbf{q}_3)^{2}(2\mathbf{p}_3^{\prime}\mathbf{q}_2^{\prime}-2\mathbf{p}_2^{\prime}\mathbf{q}_3^{\prime}-\mathbf{q}_3\mathbf{p}_2^{\prime\prime}+\mathbf{q}_2\mathbf{p}_3^{\prime\prime}+\mathbf{p}_3\mathbf{q}_2^{\prime\prime}-\mathbf{p}_2\mathbf{q}_3^{\prime\prime})}{4\mathbf{A}_1^{\frac{5}{2}}}
    \end{aligned}
\end{align}
\end{tiny}
\begin{tiny}
\begin{align}
    \begin{aligned}
        y^{\prime\prime}=& \frac{-8\mathbf{A}_1(\mathbf{q}_3\mathbf{p}^{\prime}_1+\mathbf{p}_1\mathbf{q}^{\prime}_3-\mathbf{q}_1\mathbf{p}^{\prime}_3-\mathbf{p}_3\mathbf{q}^{\prime}_1)\left[ (\mathbf{p}_2\mathbf{q}_1-\mathbf{p}_1\mathbf{q}_2)(\mathbf{q}_1\mathbf{p}_2^{\prime}+\mathbf{p}_2\mathbf{q}_1^{\prime}-\mathbf{q}_2\mathbf{p}_1^{\prime}-\mathbf{p}_1\mathbf{q}_2^{\prime})+(\mathbf{p}_3\mathbf{q}_1-\mathbf{p}_1\mathbf{q}_3)(\mathbf{q}_1\mathbf{p}_3^{\prime}+\mathbf{p}_3\mathbf{q}_1^{\prime}-\mathbf{q}_3\mathbf{p}_1^{\prime}-\mathbf{p}_1\mathbf{q}_3^{\prime})+(\mathbf{p}_3\mathbf{q}_2-\mathbf{p}_2\mathbf{q}_3)(\mathbf{q}_2\mathbf{p}_3^{\prime}+\mathbf{p}_3\mathbf{q}_2^{\prime}-\mathbf{q}_3\mathbf{p}_2^{\prime}-\mathbf{p}_2\mathbf{q}_3^{\prime}) \right] }{4\mathbf{A}_1^{\frac{5}{2}}} \\
        & -\frac{4\mathbf{A}_1^{2}(2\mathbf{p}_3^{\prime}\mathbf{q}_1^{\prime}-2\mathbf{q}_3^{\prime}\mathbf{p}_1^{\prime}-\mathbf{q}_3\mathbf{p}_1^{\prime\prime}+\mathbf{q}_1\mathbf{p}_3^{\prime\prime}+\mathbf{p}_3\mathbf{q}_1^{\prime\prime}-\mathbf{p}_1\mathbf{q}_3^{\prime\prime})}{4\mathbf{A}_1^{\frac{5}{2}}} \\
        & -\frac{12(\mathbf{p}_3\mathbf{q}_1-\mathbf{p}_1\mathbf{q}_3)\left[ (\mathbf{p}_2\mathbf{q}_1-\mathbf{p}_1\mathbf{q}_2)(\mathbf{q}_1\mathbf{p}_2^{\prime}+\mathbf{p}_2\mathbf{q}_1^{\prime}-\mathbf{q}_2\mathbf{p}_1^{\prime}-\mathbf{p}_1\mathbf{q}_2^{\prime})+(\mathbf{p}_3\mathbf{q}_1-\mathbf{p}_1\mathbf{q}_3)(\mathbf{q}_1\mathbf{p}_3^{\prime}+\mathbf{p}_3\mathbf{q}_1^{\prime}-\mathbf{q}_3\mathbf{p}_1^{\prime}-\mathbf{p}_1\mathbf{q}_3^{\prime})+(\mathbf{p}_3\mathbf{q}_2-\mathbf{p}_2\mathbf{q}_3)(\mathbf{q}_2\mathbf{p}_3^{\prime}+\mathbf{p}_3\mathbf{q}_2^{\prime}-\mathbf{q}_3\mathbf{p}_2^{\prime}-\mathbf{p}_2\mathbf{q}_3^{\prime}) \right ]^{2}}{4\mathbf{A}_1^{\frac{5}{2}}} \\
        & +\frac{4\mathbf{A}_1(\mathbf{p}_3\mathbf{q}_1-\mathbf{p}_1\mathbf{q}_3)\left[(\mathbf{q}_2\mathbf{p}_1^{\prime}+\mathbf{p}_1\mathbf{q}_2^{\prime}-\mathbf{q}_1\mathbf{p}_2^{\prime}-\mathbf{p}_2\mathbf{q}_1^{\prime})^{2}+(\mathbf{q}_3\mathbf{p}_1^{\prime}+\mathbf{p}_1\mathbf{q}_3^{\prime}-\mathbf{q}_1\mathbf{p}_3^{\prime}-\mathbf{p}_3\mathbf{q}_1^{\prime})^{2}+(\mathbf{q}_3\mathbf{p}_2^{\prime}+\mathbf{p}_2\mathbf{q}_3^{\prime}-\mathbf{q}_2\mathbf{p}_3^{\prime}-\mathbf{p}_3\mathbf{q}_2^{\prime})^{2}\right ]}{4\mathbf{A}_1^{\frac{5}{2}}} \\
        & +\frac{4\mathbf{A}_1(\mathbf{p}_3\mathbf{q}_1-\mathbf{p}_1\mathbf{q}_3)\left[(\mathbf{p}_2\mathbf{q}_1-\mathbf{p}_1\mathbf{q}_2)(2\mathbf{p}_2^{\prime}\mathbf{q}_1^{\prime}-2\mathbf{p}_1^{\prime}\mathbf{q}_2^{\prime}-\mathbf{q}_2\mathbf{p}_1^{\prime\prime}+\mathbf{q}_1\mathbf{p}_2^{\prime\prime}+\mathbf{p}_2\mathbf{q}_1^{\prime\prime}-\mathbf{p}_1\mathbf{q}_2^{\prime\prime})+(\mathbf{p}_3\mathbf{q}_1-\mathbf{p}_1\mathbf{q}_3)(2\mathbf{p}_3^{\prime}\mathbf{q}_1^{\prime}-2\mathbf{p}_1^{\prime}\mathbf{q}_3^{\prime}-\mathbf{q}_3\mathbf{p}_1^{\prime\prime}+\mathbf{q}_1\mathbf{p}_3^{\prime\prime}+\mathbf{p}_3\mathbf{q}_1^{\prime\prime}-\mathbf{p}_1\mathbf{q}_3^{\prime\prime})\right]}{4\mathbf{A}_1^{\frac{5}{2}}} \\
        & +\frac{4\mathbf{A}_1(\mathbf{p}_3\mathbf{q}_1-\mathbf{p}_1\mathbf{q}_3)^{2}(2\mathbf{p}_3^{\prime}\mathbf{q}_2^{\prime}-2\mathbf{p}_2^{\prime}\mathbf{q}_3^{\prime}-\mathbf{q}_3\mathbf{p}_2^{\prime\prime}+\mathbf{q}_2\mathbf{p}_3^{\prime\prime}+\mathbf{p}_3\mathbf{q}_2^{\prime\prime}-\mathbf{p}_2\mathbf{q}_3^{\prime\prime})}{4\mathbf{A}_1^{\frac{5}{2}}}
    \end{aligned}
\end{align}
\end{tiny}
\begin{tiny}
\begin{align}
    \begin{aligned}
        z^{\prime\prime}=& \frac{8\mathbf{A}_1(\mathbf{q}_2\mathbf{p}^{\prime}_1+\mathbf{p}_1\mathbf{q}^{\prime}_2-\mathbf{q}_1\mathbf{p}^{\prime}_2-\mathbf{p}_2\mathbf{q}^{\prime}_1)\left[ (\mathbf{p}_2\mathbf{q}_1-\mathbf{p}_1\mathbf{q}_2)(\mathbf{q}_1\mathbf{p}_2^{\prime}+\mathbf{p}_2\mathbf{q}_1^{\prime}-\mathbf{q}_2\mathbf{p}_1^{\prime}-\mathbf{p}_1\mathbf{q}_2^{\prime})+(\mathbf{p}_3\mathbf{q}_1-\mathbf{p}_1\mathbf{q}_3)(\mathbf{q}_1\mathbf{p}_3^{\prime}+\mathbf{p}_3\mathbf{q}_1^{\prime}-\mathbf{q}_3\mathbf{p}_1^{\prime}-\mathbf{p}_1\mathbf{q}_3^{\prime})+(\mathbf{p}_3\mathbf{q}_2-\mathbf{p}_2\mathbf{q}_3)(\mathbf{q}_2\mathbf{p}_3^{\prime}+\mathbf{p}_3\mathbf{q}_2^{\prime}-\mathbf{q}_3\mathbf{p}_2^{\prime}-\mathbf{p}_2\mathbf{q}_3^{\prime}) \right] }{4\mathbf{A}_1^{\frac{5}{2}}} \\
        & +\frac{4\mathbf{A}_1^{2}(2\mathbf{p}_2^{\prime}\mathbf{q}_1^{\prime}-2\mathbf{q}_2^{\prime}\mathbf{p}_1^{\prime}-\mathbf{q}_2\mathbf{p}_1^{\prime\prime}+\mathbf{q}_1\mathbf{p}_2^{\prime\prime}+\mathbf{p}_2\mathbf{q}_1^{\prime\prime}-\mathbf{p}_1\mathbf{q}_2^{\prime\prime})}{4\mathbf{A}_1^{\frac{5}{2}}} \\
        & +\frac{12(\mathbf{p}_2\mathbf{q}_1-\mathbf{p}_1\mathbf{q}_2)\left[ (\mathbf{p}_2\mathbf{q}_1-\mathbf{p}_1\mathbf{q}_2)(\mathbf{q}_1\mathbf{p}_2^{\prime}+\mathbf{p}_2\mathbf{q}_1^{\prime}-\mathbf{q}_2\mathbf{p}_1^{\prime}-\mathbf{p}_1\mathbf{q}_2^{\prime})+(\mathbf{p}_3\mathbf{q}_1-\mathbf{p}_1\mathbf{q}_3)(\mathbf{q}_1\mathbf{p}_3^{\prime}+\mathbf{p}_3\mathbf{q}_1^{\prime}-\mathbf{q}_3\mathbf{p}_1^{\prime}-\mathbf{p}_1\mathbf{q}_3^{\prime})+(\mathbf{p}_3\mathbf{q}_2-\mathbf{p}_2\mathbf{q}_3)(\mathbf{q}_2\mathbf{p}_3^{\prime}+\mathbf{p}_3\mathbf{q}_2^{\prime}-\mathbf{q}_3\mathbf{p}_2^{\prime}-\mathbf{p}_2\mathbf{q}_3^{\prime}) \right ]^{2}}{4\mathbf{A}_1^{\frac{5}{2}}} \\
        & -\frac{4\mathbf{A}_1(\mathbf{p}_2\mathbf{q}_1-\mathbf{p}_1\mathbf{q}_2)\left[(\mathbf{q}_2\mathbf{p}_1^{\prime}+\mathbf{p}_1\mathbf{q}_2^{\prime}-\mathbf{q}_1\mathbf{p}_2^{\prime}-\mathbf{p}_2\mathbf{q}_1^{\prime})^{2}+(\mathbf{q}_3\mathbf{p}_1^{\prime}+\mathbf{p}_1\mathbf{q}_3^{\prime}-\mathbf{q}_1\mathbf{p}_3^{\prime}-\mathbf{p}_3\mathbf{q}_1^{\prime})^{2}+(\mathbf{q}_3\mathbf{p}_2^{\prime}+\mathbf{p}_2\mathbf{q}_3^{\prime}-\mathbf{q}_2\mathbf{p}_3^{\prime}-\mathbf{p}_3\mathbf{q}_2^{\prime})^{2}\right ]}{4\mathbf{A}_1^{\frac{5}{2}}} \\
        & -\frac{4\mathbf{A}_1(\mathbf{p}_2\mathbf{q}_1-\mathbf{p}_1\mathbf{q}_2)\left[(\mathbf{p}_2\mathbf{q}_1-\mathbf{p}_1\mathbf{q}_2)(2\mathbf{p}_2^{\prime}\mathbf{q}_1^{\prime}-2\mathbf{p}_1^{\prime}\mathbf{q}_2^{\prime}-\mathbf{q}_2\mathbf{p}_1^{\prime\prime}+\mathbf{q}_1\mathbf{p}_2^{\prime\prime}+\mathbf{p}_2\mathbf{q}_1^{\prime\prime}-\mathbf{p}_1\mathbf{q}_2^{\prime\prime})+(\mathbf{p}_3\mathbf{q}_1-\mathbf{p}_1\mathbf{q}_3)(2\mathbf{p}_3^{\prime}\mathbf{q}_1^{\prime}-2\mathbf{p}_1^{\prime}\mathbf{q}_3^{\prime}-\mathbf{q}_3\mathbf{p}_1^{\prime\prime}+\mathbf{q}_1\mathbf{p}_3^{\prime\prime}+\mathbf{p}_3\mathbf{q}_1^{\prime\prime}-\mathbf{p}_1\mathbf{q}_3^{\prime\prime})\right]}{4\mathbf{A}_1^{\frac{5}{2}}} \\
        & -\frac{4\mathbf{A}_1(\mathbf{p}_2\mathbf{q}_1-\mathbf{p}_1\mathbf{q}_2)^{2}(2\mathbf{p}_3^{\prime}\mathbf{q}_2^{\prime}-2\mathbf{p}_2^{\prime}\mathbf{q}_3^{\prime}-\mathbf{q}_3\mathbf{p}_2^{\prime\prime}+\mathbf{q}_2\mathbf{p}_3^{\prime\prime}+\mathbf{p}_3\mathbf{q}_2^{\prime\prime}-\mathbf{p}_2\mathbf{q}_3^{\prime\prime})}{4\mathbf{A}_1^{\frac{5}{2}}}
    \end{aligned}
\end{align}
\end{tiny}
where $ \mathbf{A}_1$ is derived as:
\begin{small}
\begin{equation*}
    \mathbf{A}_1=(\mathbf{p}_2\mathbf{q}_1-\mathbf{p}_1\mathbf{q}_2)^{2}+(\mathbf{p}_3\mathbf{q}_1-\mathbf{p}_1\mathbf{q}_3)^{2}+(\mathbf{p}_3\mathbf{q}_2-\mathbf{p}_2\mathbf{q}_3)^{2}
\end{equation*}
\end{small}

Numerical methods are applied to calculate the maximum of the second-order derivatives given above.

\bibliography{sample}


\end{document}